\documentclass[11pt]{article}
\usepackage[latin1]{inputenc}
\usepackage{amsmath, amsfonts, amssymb}
\usepackage[squaren]{SIunits}
\usepackage{parskip}
\usepackage{indentfirst}
\usepackage[hidelinks]{hyperref}
\usepackage{pstool}
\usepackage[font={small},textfont=small,labelfont=bf]{caption}
\usepackage[figurename=Supporting Figure, tablename=Supporting Table]{caption}
\usepackage[a4paper,top=2cm, bottom=2cm, left=2cm, right=2cm]{geometry}
\usepackage{placeins}
\usepackage{pdfpages}

\newcommand{\e}[1]{\cdot10^{#1}}
\newcommand{\dx}{\mathrm{d}x}

\newcommand{\includeordonotincludegraphics}[2]{\resizebox{\textwidth}{!}{\includegraphics[#1]{#2_smallFromEps}}}
\newcommand{\fpf}[1]{({\bf\MakeUppercase{#1}})}
\newcommand{\fpl}[1]{{\bf\MakeUppercase{#1}}}
\newcommand{\fpt}[1]{\MakeUppercase{#1}}

\newcommand{\TromborgPRLrefnumber}{11}
\newcommand{\AmundsenTribolLettrefnumber}{28}
\newcommand{\Highlightchangesornot}{\color{black}}

\providecommand{\abs}[1]{\lvert#1\rvert}
\begin{document}

% %%%%%%%%%%%%%%%%%%%%%%%%%%%%%%%%%%%%%%%%%%%%%%%%%%%%%%%%%%%%%%%%%%%%%%%%%%%%%%%%%%%%%%%%%%%%%%%%%%%%%%%%%%%%%%%%%%%%%%%
% The main manuscript and the captions were produced in Word. Merge the pdf output from word with the supplementary information, which is in this file.
% %%%%%%%%%%%%%%%%%%%%%%%%%%%%%%%%%%%%%%%%%%%%%%%%%%%%%%%%%%%%%%%%%%%%%%%%%%%%%%%%%%%%%%%%%%%%%%%%%%%%%%%%%%%%%%%%%%%%%%%


# Slow slip and the transition from fast to slow fronts in the rupture of frictional interfaces

J. K. Trømborg[a], H. A. Sveinsson[a], J. Scheibert[b], K. Thøgersen[a], D. S. Amundsen[c], A. Malthe-Sørenssen[a,1]

    a. Department of Physics, University of Oslo, Oslo, NO 0316

    b. Laboratoire de Tribologie et Dynamique des Systèmes, CNRS, Ecole Centrale de Lyon, Ecully, FR 69134

    c. Astrophysics Group, University of Exeter, Exeter, GB EX4 4QL

    1. malthe@fys.uio.no, +47 22856477

**The failure of the population of micro-junctions forming the frictional interface between two solids is central to fields ranging from biomechanics to seismology. This failure is mediated by the propagation along the interface of various types of rupture fronts, covering a wide range of velocities. Among them are so-called slow fronts, which are recently discovered fronts much slower than the materials' sound speeds. Despite intense modelling activity, the mechanisms underlying slow fronts remain elusive. Here, we introduce a multi-scale model capable of reproducing both the transition from fast to slow fronts in a single rupture event and the short-time slip dynamics observed in recent experiments. We identify slow slip immediately following the arrest of a fast front as a phenomenon sufficient for the front to propagate further at a much slower pace. Whether slow fronts are actually observed is controlled both by the interfacial stresses and by the width of the local distribution of forces among micro-junctions. Our results show that slow fronts are qualitatively different from faster fronts. Since the transition from fast to slow fronts is potentially as generic as slow slip, we anticipate that it might occur in the wide range of systems in which slow slip has been reported, including seismic faults.**

**Significance statement**

Relative motion between solids in frictional contact is known to start progressively, with a slipping zone growing along the interface. The propagation of the front separating the stuck and slipping zones is usually very fast, but it can also slow down considerably. This transition is not yet understood. Using a multi-scale model, we demonstrate how the transition to slow fronts is the direct consequence of slow slip motion at the interface, and that slow fronts are qualitatively different from faster fronts. Since the transition to slow fronts is allowed by slow slip, we expect that it can occur in the wide range of systems in which slow slip has been reported, from engineering to seismology.

**Introduction**

The rupture of frictional interfaces is a central mechanism in many processes including snow slab avalanches, human object grasping and earthquake dynamics (1). Rupture occurs through the propagation of a crack-like micro-slip front – the rupture front – across the interface. This front represents the moving boundary between a stick region and a slipping region that coexist within the contact. In so-called partial-slip situations, fronts propagate quasi-statically at a pace controlled by the external loading, as studied in mechanical engineering for decades (2, 3). Recently, fast cameras enabled the observation of much faster fronts, which are classified into three types: supershear fronts faster than the material's shear wave speed $c_s$, sub-Rayleigh fronts propagating at velocities close to $c_s$ and slow fronts much slower than $c_s$ (4-8). Whereas the two first types have been predicted theoretically, the physical mechanisms underlying slow fronts are still debated.

A better understanding of slow fronts appears as a significant step towards an improved assessment of how frictional motion begins. It is also expected to shed light on the important topic of slow earthquakes, which have been increasingly reported in the last decade (1). In this context, an intense theoretical and numerical activity arose to investigate the origins and properties of rupture fronts. Two different approaches have been explored.

On the one hand, two- or three-dimensional elastodynamic models have been used to relate the macroscopic loading conditions to the stress field along the contact interface (9-13). These local stresses were indeed shown experimentally to play a role in the selection of the front type (6). However, the models were based on simple friction laws, e.g. Coulomb friction (11) or velocity-weakening friction (12, 13) and did not exhibit slow fronts. On the other hand, fronts much slower than the speed of sound were produced by one-dimensional friction models using improved local friction laws involving an intrinsic time-scale: either an aging scale (i) related to the long-time strengthening of an interface at rest (14, 15) as is classically

considered in rate-and-state theories (16, 17), or a dynamic scale (ii) related to the short-time collective dynamics of formation and rupture of a statistical number of micro-junctions during the rupture of the interface (18, 19). These two time-scales (i) and (ii) were shown, in the same experiment (20), to control the recovery of contact area after slip arrest and the transition between a fast and a slow regime for the slip motion triggered by the front passage, respectively. However, because one-dimensional models are unable to reproduce realistic stress distributions at the interface, they do not allow for quantitative comparison with experiments. Here we combine both 2D elastodynamics and time-dependent friction into a multi-scale model for rupture fronts. We demonstrate that the model simultaneously reproduces two separate, unexplained experimental observations: the transition between fast and slow front propagation during a single rupture event; and the transition from fast to slow slip motion at the interface shortly after rupture. In addition, through a more general study of the model, we show that slow slip occurring immediately after a fast front's arrest is a sufficient phenomenon for the front to propagate further as a slow front, and we identify parameters involved in front type selection.

**Model description**

The frictional stability of a system made of two solids in contact depends on the level of normal and shear stresses at the contact interface. These interfacial stresses result from the external forces applied at the boundaries of the solids, transmitted through the bulk. Slip motion will in general be triggered when the local interfacial shear stress reaches a threshold, the level of which crucially depends on the interface behavior law at the microscale.

The net contact between two solids generically consists of a large number of stress bearing micro-junctions whose nature depends on the type of interface. For rough solids, each micro-junction corresponds to a micro-contact between antagonist asperities, whereas for smoother surfaces the junctions can be solidified patches of an adsorbate layer (21). The three physical

aspects of the junction behavior that we consider to be essential are as follows. 1) A micro-junction in its pinned state can bear a shear force $f_T$, provided it remains smaller than a threshold $f_{thres}$. When $f_{thres}$ is reached, a local fracture-like event occurs, and the junction enters a slipping state. 2) In the slipping state, the micro-junction can let the interface slip, either through the micro-slipping of micro-asperities in contact or through the fluidization of an adsorbate layer. During slip, the micro-junction sustains some residual force $f_T = f_{slip}$, with $f_{slip}$ smaller than $f_{thres}$. 3) Slipping micro-junctions have a certain probability to disappear or relax. For example, a micro-contact disappears when an asperity moves away from its antagonist asperity by a typical distance equal to the mean size of micro-contacts, as classically considered for slow frictional sliding, e.g. in rate-and-state friction laws. However, another picture may arise as a consequence of the sudden release of energy when pinned junctions break. This energy will transiently heat the region around the micro-junction (20). The rise in temperature will significantly increase the rate of a thermally activated relaxation of the slipping micro-junction during the time necessary for the interface to cool down (21). The frictional consequences of such temperature rises have recently received renewed attention (see e.g. (22)), but remain poorly understood. In an attempt to include such thermal processes in our model, we recognize that they will lead to time- rather than distance-controlled relaxations, so that the shear force drop will be distributed in time. In order for the interface to continue bearing the normal forces applied to it, the micro-junctions that relax are replaced by new, pinned junctions bearing a small tangential force $f_{new}$.

The physical aspects described above have been modeled in a simple way using the following assumptions (Methods). We consider the rough frictional interface between a rigid track and a thin linear elastic slider of length $L$ and height $H$ (Fig.1A). The bulk elastodynamics of the slider are solved using a square lattice of blocks connected by internal springs (11, 23) (Fig.1B). The multi-contact nature of the interface is modeled through an array of $N_s$

tangential springs representing individual micro-junctions, attached in parallel to each interfacial block (Fig.1C) (18, 19, 21). The individual spring behavior is as follows (Fig.1F, (24)). A spring pinned to the track stretches linearly elastically as the block moves, acting with a tangential force $f_T$ on the block. When the force reaches the static friction threshold $f_{thres}$ (we neglect aging, so that $f_{thres}$ is time independent), the micro-junction ruptures and the spring becomes a slipping spring acting with a dynamic friction force[*] $f_T=f_{slip}$. After a random time $t_R$ drawn from a distribution $T(t_R)$, the slipping spring relaxes. It is replaced immediately by a pinned, unloaded spring ($f_{new}=0$) and a new cycle starts. Here we use $T(t_R)$ as a simplified way of modeling the distribution of times after which micro-junctions relax. Due to the variety and the complexity of the underlying thermal processes, we did not try to derive $T(t_R)$ for a specific situation. Rather, we chose to model $T(t_R)$ in the simplest way, as a Gaussian with average time $<t_R>$ and width $\delta t_R$. The shape of $T(t_R)$ is not crucial: we obtain qualitatively similar results with an exponential distribution. The width of $T(t_R)$ is the only source of randomness in our model and causes the interface springs of a block to evolve differently from each other.

**The model reproduces rupture front observations**
In this section we use the model to reproduce unexplained experimental observations made in polymethylmethacrylate by Fineberg and coworkers. We use the loading conditions and material constants of (4), as done in (11). We then select the parameters of the friction law in order to reproduce both the complex space-time evolution of the front velocity reported in (4) and the time-dependent interfacial slip history reported in (20).

---

[*] $f_{thres}$ and $f_{slip}$ are taken proportional to the normal force $p$ on the corresponding block (Methods). This assumption ascribes the pressure dependence of the forces on a block to the individual forces bore by a constant number $N_s$ of springs, rather than to a pressure dependent number of springs per block.

The slider is first submitted to normal load $F_N$ only. Then, a slowly increasing tangential load $F_T$ is applied on its left side, at height $h$ above the interface (Supporting Methods). Macroscopically (Fig.1D), the slider is first loaded elastically, yielding high shear stresses in the vicinity of the loading point. These growing stresses eventually trigger the slip of a first block, the motion of which increases the force on its neighbors (Fig.1E) until they also start to slip, and so on. This mechanism results in the propagation of a rupture front across the interface. The first such events arrest before reaching the contact's leading edge and correspond to so-called precursors to sliding (11, 25-28). In the following, we will only consider interface-spanning events (Fig.2), that are responsible for the large force drops in Fig.1D. We measure the front propagation speed $v_c$ as a function of position $x$ along the interface (Fig.2C) by defining rupture at the block scale, when the fraction of pinned springs drops below a threshold value (Figs.2A,B).

Figure 2A shows an event starting as a fast front ($v_c \sim c_s/3$), then turning into a slow front ($v_c \sim c_s/100$) at $x \sim 0.4L$, before turning back into a fast front at $x \sim 0.6L$. This space-time development is in excellent agreement with the experiments from which the model parameters were taken (4). In particular, the locations of the transitions between fast and slow fronts, the duration of the slow front (~1ms) and the velocity ratio between fast and slow fronts are all matching the experimental observations. Other events are fast across the whole interface (Fig.2B). We emphasize that such spatially heterogeneous dynamics arise spontaneously in our model. This success is presumably due to the two-dimensionality of the model, which allows it to reproduce interfacial stress heterogeneities arising from macroscopic sample geometry and external loading (11) (Fig.S1).

Figure 3C shows the slip experienced by an interfacial block during an event, the time-evolution of which exhibits two distinct regimes. A fast slip regime is followed by a roughly linear slow slip regime, in good agreement with the slip history reported in (20). Notably,

both the ratios of slip distances and of slip velocities experienced in either regime are matching the experimental observations.

The simultaneous agreement found with two independent measurements – reproducing the fronts' spatio-temporal dynamics at the slider scale and the slip dynamics at the block scale – indicates that our model can provide insight into the set of experiments reported by Fineberg and coworkers. Because the basic ingredients of the model represent general features of frictional interfaces, we expect the model to be applicable to a larger class of systems.

**Relationship between slow slip and the transition to slow fronts in the model**

From this section on we change the scope from reproducing particular experiments to performing a systematic study of the model. We begin by varying select model parameters to unravel the mechanisms underlying slow fronts.

Figure 3C (reference curve) shows the slip history of a block during an event with a slow front (Fig.3A); it has both fast and slow slip regimes. Fast slip initiates with the passage of the rupture front and is independent of $<t_R>$ (the three curves in Fig.3C overlap during fast slip). It is followed by a short period of arrest. Then, slow slip originates from the following mechanism: For each spring leaving the slipping state after $t_R$ the friction force on the block is reduced. As a result the block moves a small distance, driven by the forces from the neighboring blocks. Slow slip results from the succession of such incremental block movements. A scaling analysis of this mechanism (Supporting Equations) yields a prediction for the slow slip velocity $v_{\text{slowslip}}$, which quantitatively captures the data (inset of Fig.3C).

The fast-slow-fast front in (A) becomes an arrested front (B) when we restart the simulation in the same state, but with slow slip turned off. We do this by setting $f_{\text{new}} = f_{\text{slip}}$ so that springs leaving the slipping state no longer relax, but return to the pinned state bearing the same force as during slip. The front in (B) arrests where the fast-to-slow transition used to take place.

These results support the following scenario: Slow fronts are fronts that would arrest in the absence of slow slip, but are pushed further by the increasing loading due to slow slip behind the front tip.

To show that slow slip occurring immediately after front arrest is a sufficient condition for the transitions to slow fronts to be allowed, we consider the following simplified model. We leave microscale dynamics out by replacing the ensemble of springs connecting each block to the track with a single spring: The spring breaks when reaching the force $N_s f_{thres}$, it slides with a force $N_s f_{slip}$ and it returns to the pinned state still bearing the force $N_s f_{slip}$ when the block velocity vanishes. Slow slip is introduced directly as a small velocity applied to the spring attachment point (Supporting Methods). The transition from fast to slow fronts is indeed observed in this simplified model (Fig.S2), and disappears if slow slip is turned off.

To characterize the relationship between slow slip velocity and slow front velocity, we consider both the complete and simplified models and use interfaces prepared as follows (Supporting Methods). We choose the normal force $p(x)$ on each block from the substrate to be uniform in space and constant in time, and the shear force profile $\tau(x)$ to have initial values $\tau_o(x)$ uniform in space except in a region used to initiate front propagation (Fig.S3). We vary different model parameters, keeping the front initiation region unchanged, and systematically measure the slow front velocity $v_{c,slow}$ as a function of $v_{slowslip}$ (Fig.3D,inset). Suitable rescaling (Supporting Equations) of $v_{slowslip}$ allows all points to collapse on a single straight line (Fig.3D). This collapse shows that $v_{c,slow}$ is proportional to $v_{slowslip}$, with the same conversion factor for both models. We conclude that even though $v_{slowslip}$ depends on the interfacial dynamics, the conversion factor does not; it mainly depends on the way stresses are transferred from the slowly slipping region to the rupture tip. Similar linear relationships between slip and front velocities were found in previous models (14, 15, 29).

## Selection of the front type in the model

We now turn to the question of front type selection. Ben-David et al. (6) showed that slow (fast) fronts correlate locally with small (high) shear to normal stress ratio $\tau_o(x)/p(x)$ just before the event. In equilibrium, $\tau$ is the sum of the spring forces $f_T$ for each block. Due to the randomness in the slip times $t_R$ the individual $f_T$ will be different. As in previous works (21, 24, 30) it is useful to define a distribution $\phi(f_T)$ of the forces in the springs attached to each block. $\phi$ is a dynamical quantity evolving with both the loading on and the motion of blocks. Note that in general $\phi$ will be different for every block.

We find that the width $\sigma$ of the distribution is as important as the stress ratio for front type selection. To demonstrate this we take event II as a reference and modify it by changing $\sigma$ for all blocks at $x>0.3L$. To isolate the effect of $\sigma$ from the effect of the stress ratio on front propagation we increase $\sigma$, but keep $\tau_o(x)$ unchanged. We do this by making each block's $\phi(f_T)$ a uniform distribution while keeping its average unchanged (Fig.S4). We then restart the simulation and observe that the front has changed from a fast-slow-fast one (Fig.4A) to a fast one across the whole interface (Fig.4B). The reason for this behavior is that increasing $\sigma$ from Fig.4A to Fig.4B weakens the interface, which enables the fast rupture to reach the leading edge without stopping and becoming a slow front. The weakening can be understood from Fig.4C, which shows the effective static friction coefficient $\mu_s^{\text{eff}}=\tau_{\max}/p$ of an interface block, with $\tau_{\max}$ the maximum friction force on the block before it starts to slip. Small (large) $\sigma$ correspond to large (small) $\tau_{\max}$ (Fig.S5), i.e. a locally stronger (weaker) interface, consistent with the fact that points corresponding to slow fronts cluster at low $\sigma$ in Fig.4C.

Figures 4A and 4B indicate that, besides $\tau_o/p$, $\sigma$ is involved in front type selection. Using interfaces prepared as in Fig.3D, we systematically vary the values of $\tau_o$ and $\sigma$ and observe which front type is selected (Fig.4D). The effect of $\tau_o$ agrees with experiments (6), but is

modulated by the effect of $\sigma$. Low $\tau_o$ and $\sigma$ lead to fronts that arrest before spanning the whole interface. Large $\tau_o$ and $\sigma$ yield fast fronts. Global events containing a slow part are found in a region of intermediate $\tau_o$ and $\sigma$. These results are found qualitatively robust not only against changes in sample geometry or energy stored in the nucleation region, but also against changes in the individual behavior of the interfacial springs (Fig.S6). We therefore expect the main features of Fig.4D to be widely relevant.

**Discussion**

The present model differs from classical rate-and-state friction laws (16, 17), which are empirical laws based on experimental results obtained for small sliding velocities (≲100μm/s), for which self-heating of the interface is negligible. In rate-and-state models the relevant time-scale is that of aging, i.e. the slow (logarithmic) recovery of contact area at rest. The relevant slip distance is the average micro-contact size, after which all junctions are renewed. Here, we focus on a different friction regime, with a μs-long fracture process, followed by large slip velocities in the fast slip regime (~100mm/s in (20)), both phenomena contributing to a significant temperature increase at the interface. Our model thus considers the relevant time-scale $t_R$ to be the one of thermally activated relaxations during the period needed for the interface to slowly slip and stop (we neglect aging). The corresponding slip distance is the one allowed by junction relaxations during the same period.

In the context of rate-and-state friction, it was recently suggested that slow slip velocity corresponds to the sliding speed at which the interface's steady state friction laws reaches a minimum (14, 15, 31). In our model, however, slow slip has a completely different origin, related to the intrinsic relaxation dynamics of the interface after a fast slip period (Fig.3C, Supporting Equations). As a matter of fact, the steady-state friction law that emerges from the micro-junction dynamics at the block scale is purely monotonic and decreasing (Fig.1G, (24)).

To what extent the slow fronts related to both types of slow slip share similar properties remains to be investigated.

Up to now, the term slow front has been used in the literature to name any front propagating at least one order of magnitude slower than the material's wave speeds. Our results allow for a tentative definition of the slow fronts first observed in (4). They are fronts that would arrest in the absence of slow slip, but are pushed further by the increasing loading due to slow slip behind the front tip. Importantly, the motion results from the intrinsic relaxation dynamics of the interface and not from an external driving. A direct consequence is shown in Fig.5A, where the increase in tangential load is stopped during the propagation of the slow front of Fig.2. We observe that the slow front is not affected and propagates exactly like in the original event, which indicates that the front is truly dynamic. This contrasts with the behavior of the event shown in Fig.5B, where the loading is applied to the bottom-left-most block. Extremely slow fronts are then observed, similar to the ones reported in e.g. (15, 18, 19, 32), which propagate over the time-scale of the interval between two dynamic events. These fronts stop when the driving stops, indicating that they are of a different, quasi-static nature.

This distinction sheds light on the difference between our model and recent models for the onset of sliding of extended interfaces. Urbakh and coworkers (18, 19) have used 1D spring-block models with an interfacial behavior also based on micro-junctions having two possible states. However they used $f_{\text{slip}}=0$, which prevented slow slip to occur. As a consequence, they observed fast and quasi-static fronts, but did not observe dynamic slow fronts like those reported here. Bouchbinder and coworkers (14, 15, 32) have developed an improved rate-and-state law and used it in 1D. They consider the classical aging time scale whereas we consider a dynamic healing time scale relevant during the short-lived rupture-induced temperature rise of the interface. They observed the transition from quasi-static to fast dynamic rupture, but no fast to slow dynamic front transition. Other works considered 2D models, but with velocity-

weakening local friction laws (12, 13). While they observed fast and quasi-static fronts, they did not report any transition from fast to slow dynamic fronts.

Our results make a direct bridge between the separate observations of a transition from fast to slow fronts and of slow slip in recent experiments (4, 20). As slow slip was observed in many systems from geoscience (1, 33, 34) to materials science (20, 35), we expect the transition to slow fronts to be possible in these systems too. The physical process underlying slow slip, e.g. thermal softening in glassy polymers (20), thermal creep in paper (35) and dilatant strengthening in subduction zones (36) differs from system to system (22). The present model does not aim at modeling one particular process. Rather, through the time distribution $T(t_R)$, we introduce the minimal generic ingredient necessary to yield a slow interfacial slip motion.

Using the simplified model, we demonstrated that slow fronts can in principle exist even if only one micro-junction is kept per block, i.e. the force distribution $\phi$ has width $\sigma=0$, provided slow slip is introduced in another way. However, we believe that virtually all frictional interfaces are of the multi-junction type at the relevant scale, e.g. micro-contact or molecular scale. Due to random physical properties or stochastic pinning/relaxation events combined with the previous sliding dynamics there will always be some disorder in the forces bore by the junctions, as recognized in various friction models (e.g. (18, 21, 24, 30)). The disorder was shown to control e.g. the steady sliding friction force and the transition between smooth sliding and stick-slip regimes (30). Here we showed that the force distribution is also a key parameter in front type selection. It appears as a state parameter that, by controlling the effective friction threshold $\mu_s^{\text{eff}}$ at the block scale (24), is able to affect interfacial rupture in a way analogous to but different from the usual age state used in rate-and-state friction laws.

We believe that all our results are relevant to a large class of systems with a random population of stress-bearing entities having both a threshold like rupture behaviour and a

time-dependent healing process, like in nanoscience (37), polymer science (38) and seismology (39).

**Methods**

Parameters are in Supporting Table S1. Spring $j$ of block $i$ has strength $f_{thres=}\mu_s p_i/N_s$ and stiffness $k_{ij} \propto \sqrt{p_i/N_s}$. The force on block $i$ from spring $j$ is $f_{Tij} = k_{ij}(x_{ij} - x_i)$, where $x_{ij}$ is the attachment point of the spring to the track and $x_i$ is the block's position. While slipping, the spring trails the block with $x_{ij}$ adjusted in every time step to ensure $|f_{Tij}| = f_{slip} = \mu_d p_i/N_s$. Note that if the block motion reverses, $x_{ij}$ stops changing and remains fixed while $|f_{Tij}| \leq f_{slip}$. The probability distribution $T(t_R)$ is based on a Gaussian $T_G(t_R) = \frac{1}{\sqrt{2\pi}\delta t_R}\exp\left(-\frac{(t_R-\bar{t}_R)^2}{2\delta t_R^2}\right)$, modified so that negative lifetimes are forbidden: $T(t_R) = T_G(t_R) + T_G(-t_R), t_R \in [0, \infty]$. With the parameter values we use, $<t_R> \cong \bar{t}_R$. In the simulations behind Fig. 3C and 3D, the ratio $\delta t_R/\bar{t}_R$ is maintained. Details on initialization and non-frictional boundary conditions are in Supporting Methods. The motion is found using a Velocity Verlet scheme with $\Delta t = 2 \cdot 10^{-7}$s.

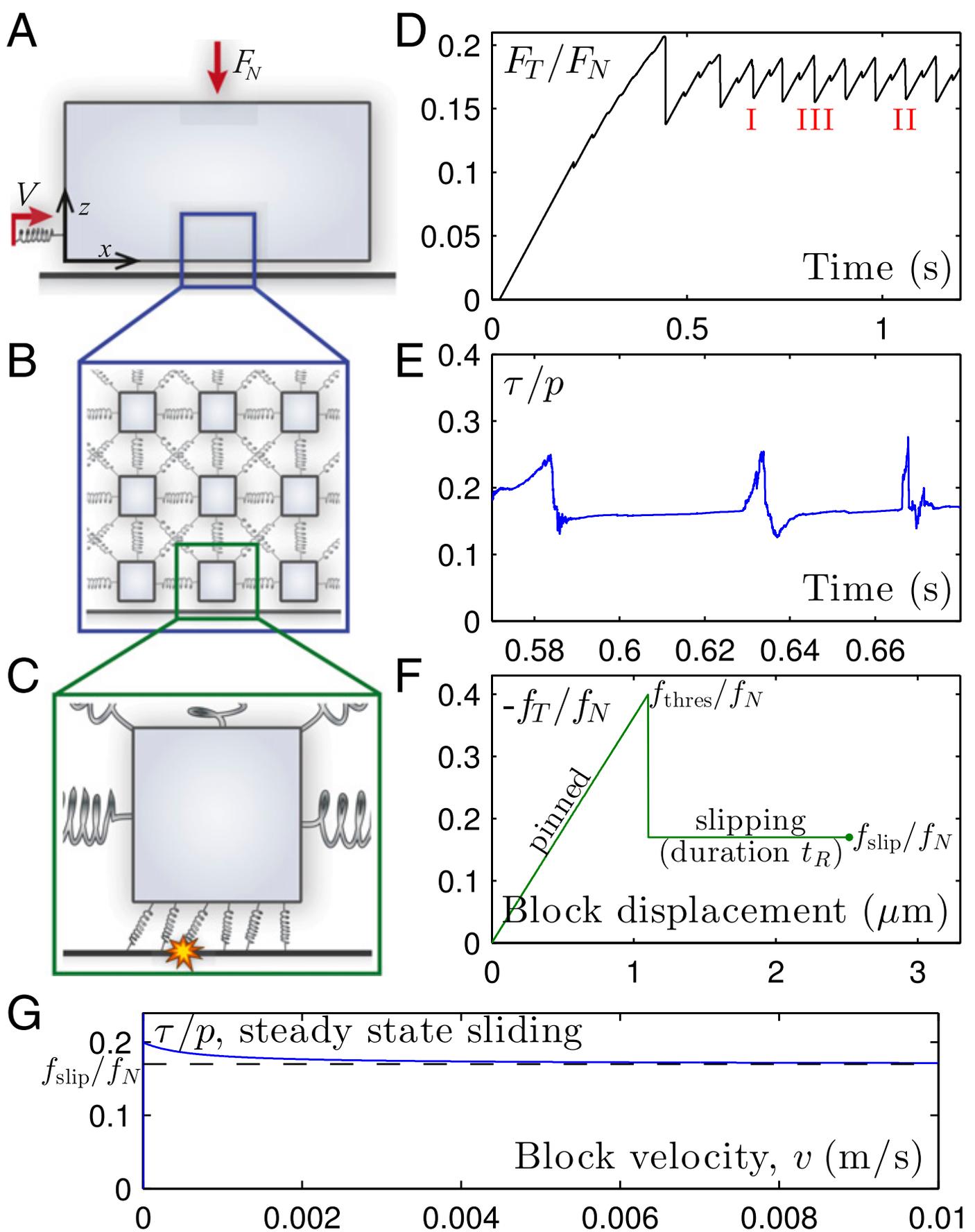

**Fig. 1.** Sketch and behavior of the multiscale model. (*A*) Slider and external loading conditions. (*B*) Spring-block network modeling elastodynamics. (*C*) Surface springs modeling friction on a block. (*D*) Macroscopic loading curve, the ratio $F_T/F_N$ of driving shear force to total normal force. (*E*) Mesoscopic loading curve, the ratio $\tau/p$ of shear to normal stress on a block. (*F*) Microscopic friction model for the spring loading curve, the ratio $f_T/f_N$ of friction to normal force for one spring ($f_N = p/N_s$). (*G*) Steady-state friction force on a block vs. sliding velocity.

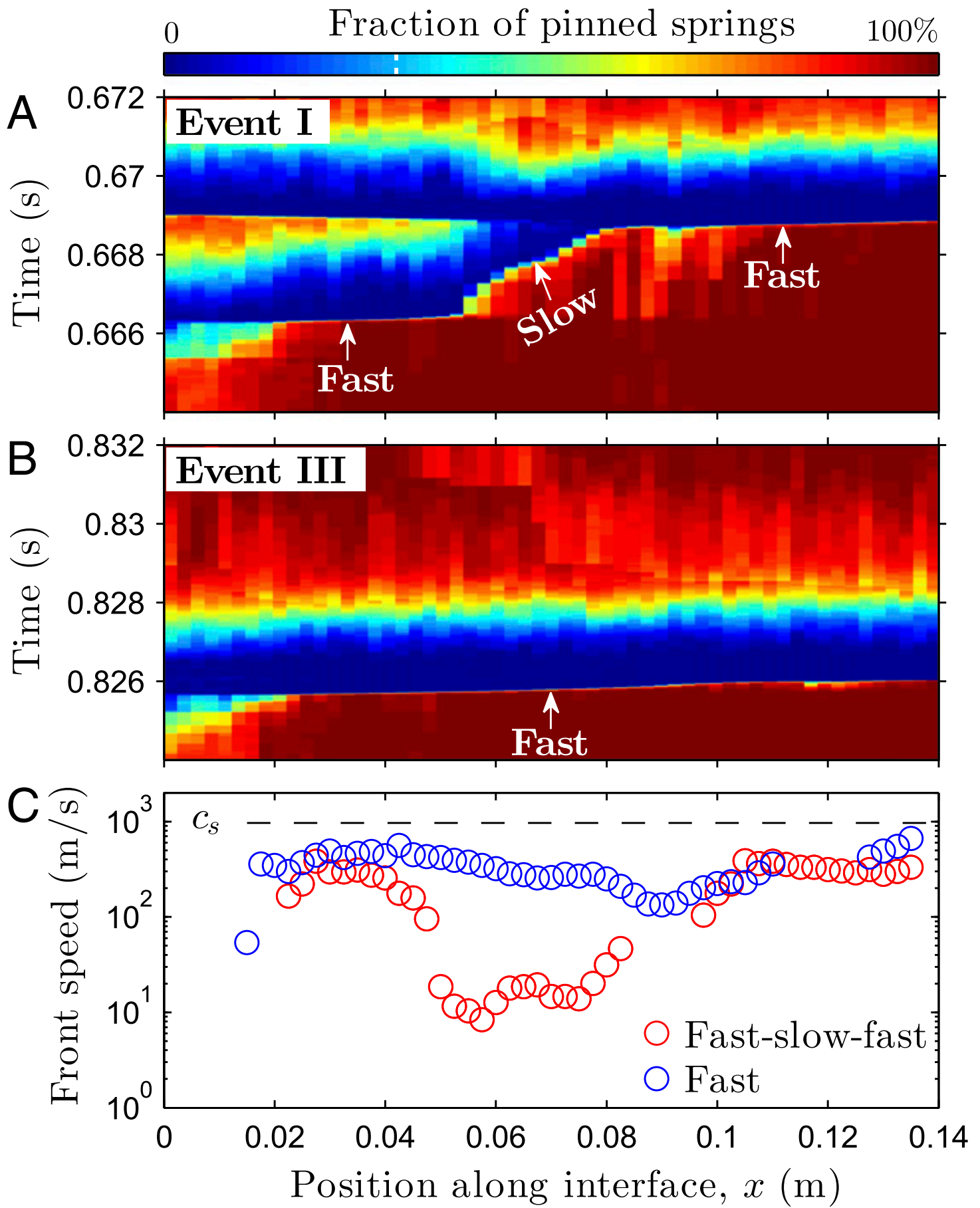

**Fig. 2.** Two interface-sized events. (*A*) A fast–slow–fast event (I in Fig. 1*D*). Spatiotemporal plot of the fraction of pinned springs. (*B*) A fast-only event (III in Fig. 1*D*) shown as in *A*. (*C*) Rupture front speed $v_c$ vs. front location for both events. Block rupture is defined to occur when 70% of interface springs have broken (white dashed line in the colorbar). Front speed is measured as the inverse slope of the rupture line (indicated by arrows in *A* and *B*) using the endpoints in a five-point-wide moving stencil.

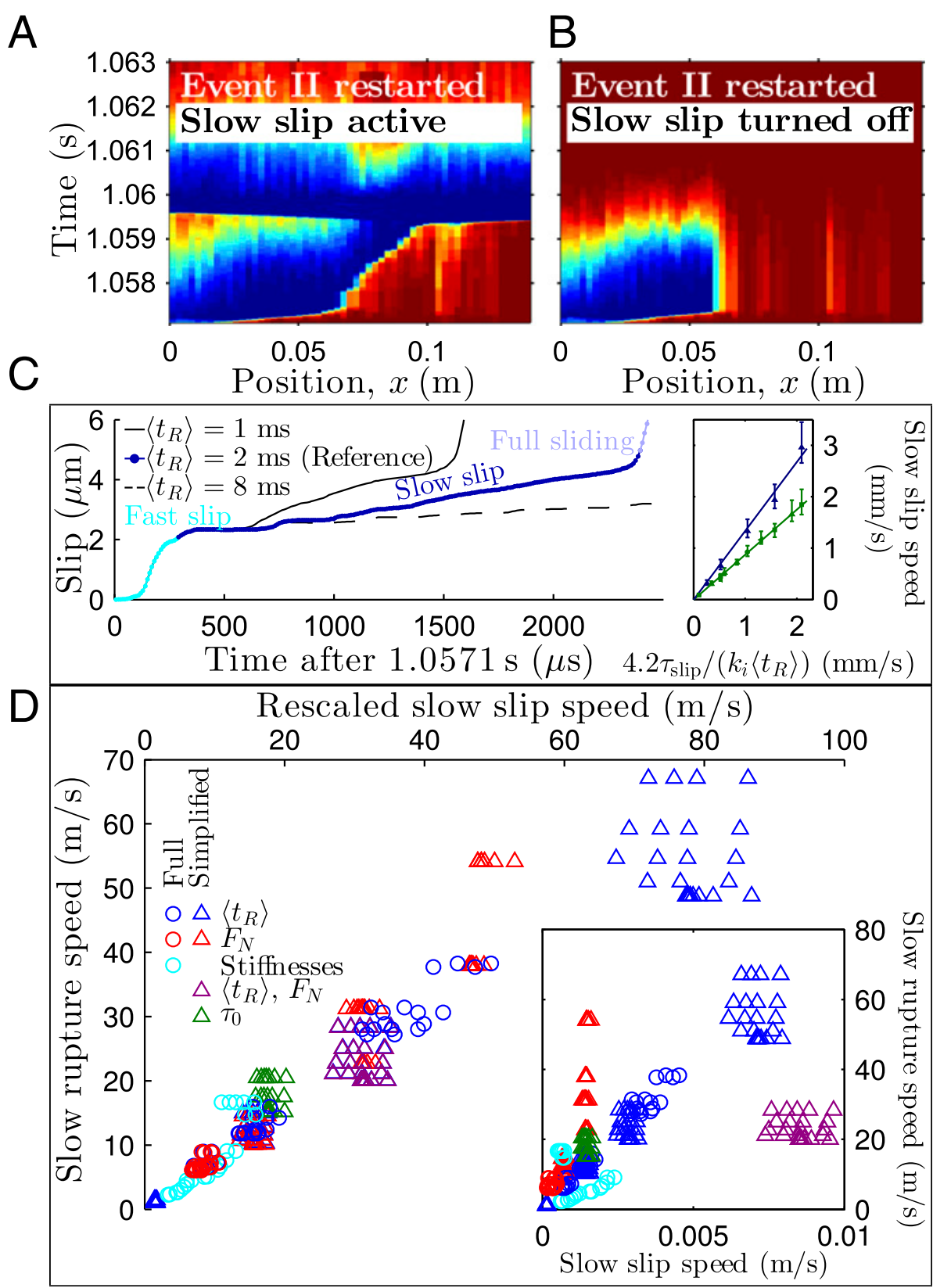

**Fig. 3.** Slow rupture is governed by slow slip. (*A* and *B*) Event II restarted at 1.0571 s with driving speed $V = 0$, shown as in Fig. 2*A*. Slow slip is either active (*A*) or turned off (*B*). (*C*) Slip profiles for block at $x = 0.34L$ under varying $\langle t_R \rangle$ (*A* uses the reference $\langle t_R \rangle$). The fast slip part is not affected. (*Inset*) Measured slow slip speed $v_{\text{slow slip}}$ matches $v_{\text{slow slip, estimate}} = 4.2\tau_{\text{slip}}/k_i\langle t_R \rangle$ (*SI Equations*). Blue represents data from full simulation and green the data for prepared homogeneous interfaces. $k_i$ is the block–interface stiffness, $\tau_{\text{slip}} = N_s f_{\text{slip}}$. Error bars represent minimum and maximum values. Markers represent the mean (*SI Equations*). The line represents linear fit to mean values, through origin. (*D*) Data collapse of slow rupture speed vs. $v_{\text{slow slip}} k_i l_0/(\tau_{\text{thres}} - \tau_0)$ along prepared homogeneous interfaces, for full and simplified models, and for wide variations in input parameters (see key within the graph). $l_0$ is a characteristic length (*SI Equations*), $\tau_{\text{thres}} = N_s f_{\text{thres}}$. (*Inset*) Unscaled data.

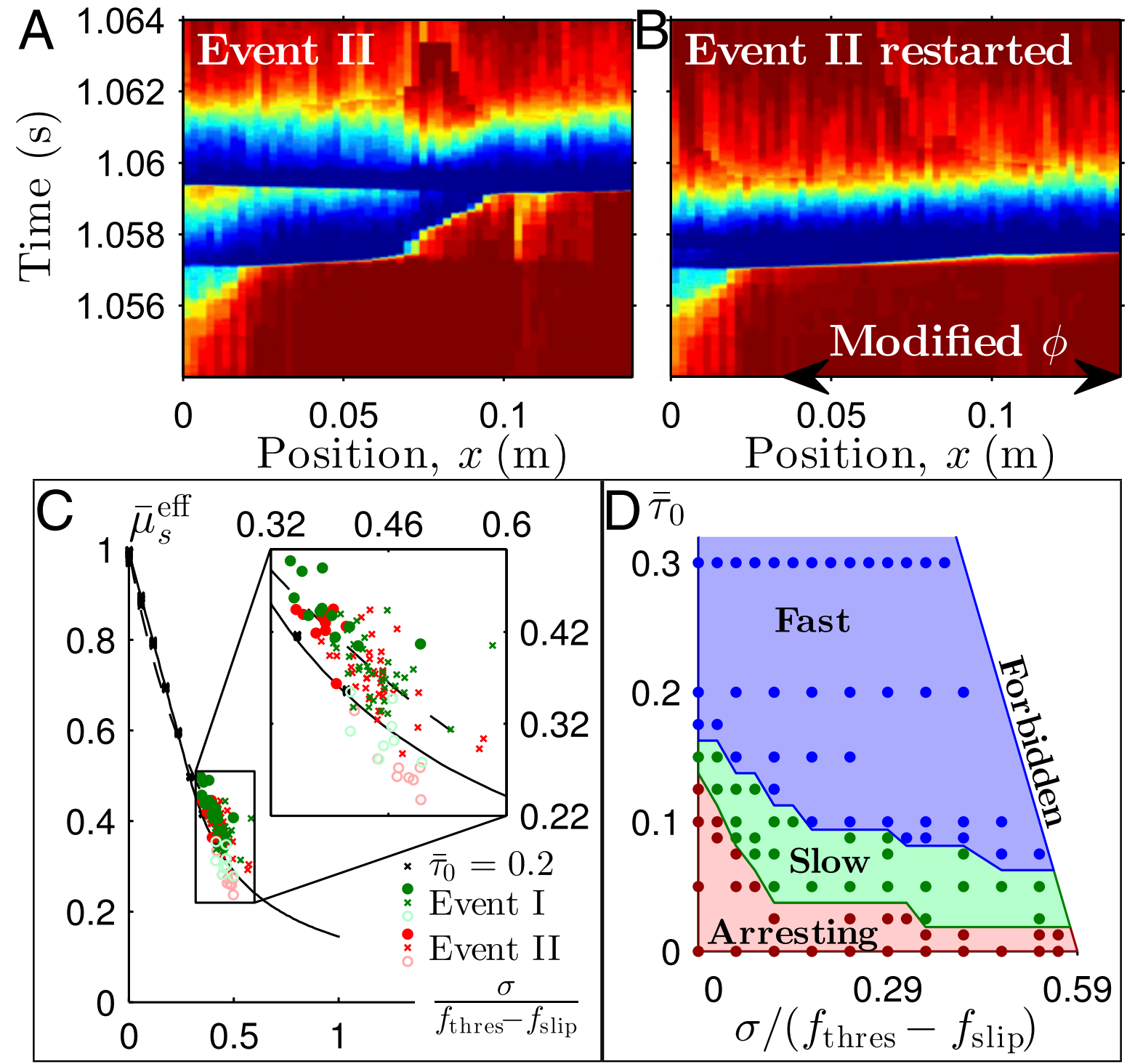

**Fig. 4.** Role of microscopic force distributions in rupture dynamics. (*A*) Event II shown as in Fig. 2*A*. (*B*) Event II restarted at 1.054 s with wider $\phi(f_T)$ in the region $x > 0.25L$ (arrow) results in a fast-only event. (*C*) $\bar{\mu}_s^{\text{eff}} = ((\tau_{\max}/p) - (f_{\text{slip}}/f_N))/((f_{\text{thres}}/f_N) - (f_{\text{slip}}/f_N))$ vs. $\tilde{\sigma} = \sigma/(f_{\text{thres}} - f_{\text{slip}})$ for all blocks in events I (green) and II (red), and for prepared events (black) with homogeneous $\bar{\tau}_0 = ((\tau_0/p) - (f_{\text{slip}}/f_N))/((f_{\text{thres}}/f_N) - (f_{\text{slip}}/f_N))$. ●, slow front; ×, fast front; ○, loading zone ($x < 2$ cm). Solid (dashed) line is for uniform (bell-shaped) $\phi(f_T)$ (Fig. S4). (*D*) Observed front type for prepared interfaces vs. both $\bar{\tau}_0$ and $\tilde{\sigma}$. Dots represents data points.

22

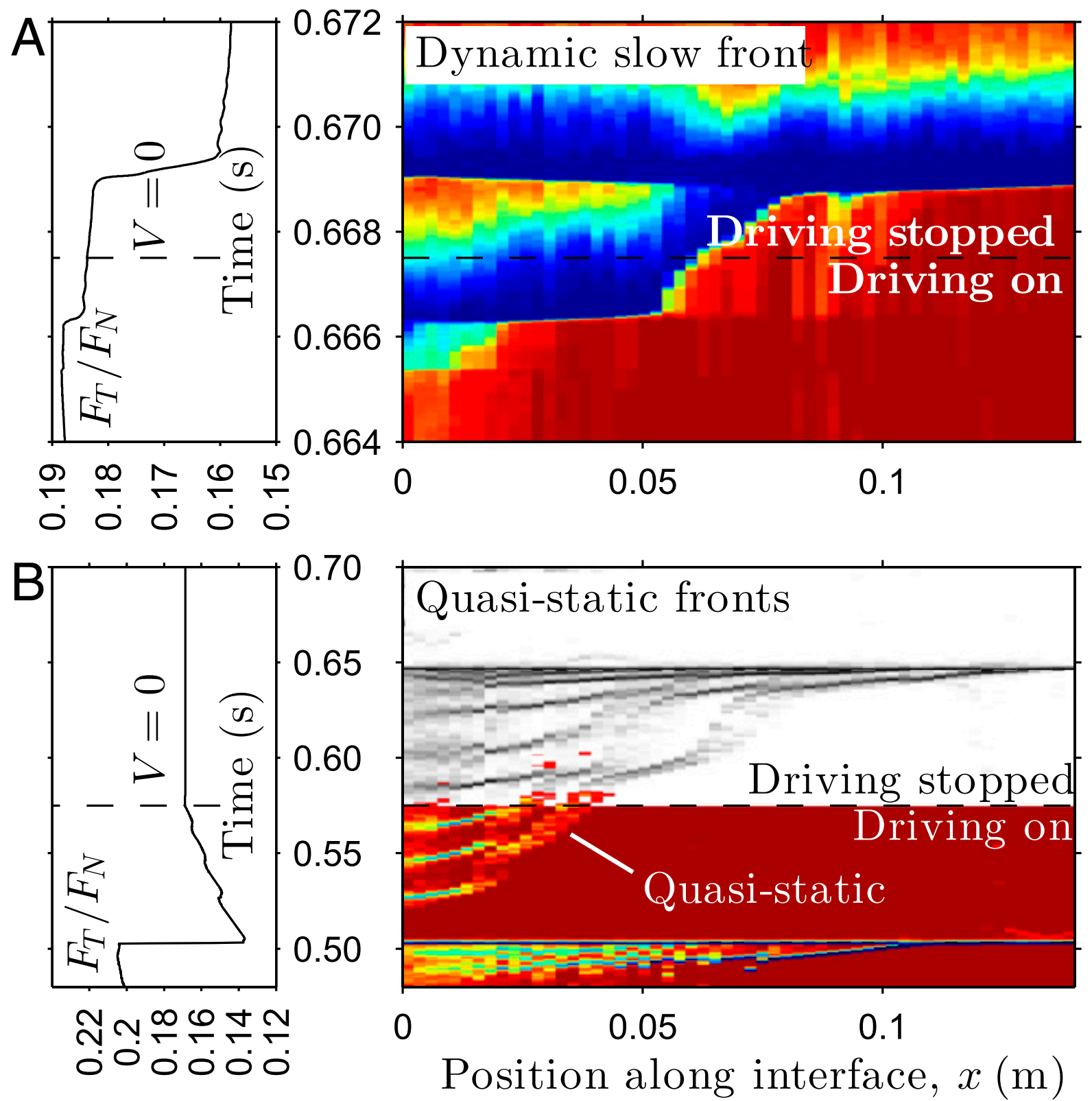

**Fig. 5.** Dynamic vs. quasistatic slow fronts. (*Upper* and *Lower Left*) Macroscopic loading curves. (*Upper* and *Lower Right*) Fraction of pinned springs shown as in Fig. 2*A*. (*A*) Event I restarted. (*B*) Data from a simulation where driving is applied at $h = 0$. Gray scale represents the original data with no change to $V$. Color represents the data in which the driving speed is set to $V = 0$ during front propagation. After setting $V = 0$, we only show color when at least 5% of a block's springs are slipping.

23



% %%%%%%%%%%%%%%%%%%%%%%%%%%%%%%%%%%%%%%%%%%%%%%%%%%%%%%%%%%%%%%%%%%%%%%%%%%%%%%%%%%%%%%%%%%%%%%%%%%%%%%%%%%%%%%%%%%%%%%%
% Supporting information
% %%%%%%%%%%%%%%%%%%%%%%%%%%%%%%%%%%%%%%%%%%%%%%%%%%%%%%%%%%%%%%%%%%%%%%%%%%%%%%%%%%%%%%%%%%%%%%%%%%%%%%%%%%%%%%%%%%%%%%%

\begin{center}
Supporting information for\\
{\huge Slow slip and the transition from fast to slow fronts in the rupture of frictional interfaces}\\
J.~K.~Trømborg, H.~A.~Sveinsson, J.~Scheibert, K.~Thøgersen, D.~S.~Amundsen, A.~Malthe-Sørenssen
\end{center}

% %%%%%%%%%%%%%%%%%%%%%%%%%%%%%%%
% Extended Data Table
% %%%%%%%%%%%%%%%%%%%%%%%%%%%%%%%
\begin{table}[b]
\centering
\caption{Model parameters. Parameters above the horizontal line are used in the same way as in Ref.~(\TromborgPRLrefnumber).\label{tab:parameters}} %Ref.12 is Trømborg et al., PRL, 2011.
\begin{tabular}{lll}\hline
Name				& Symbol	& Value				\\\hline
Slider length ($x$)		& $L$		& $\unit{140}{\milli\metre}$	\\
Slider height ($z$)		& $H$		& $\unit{75}{\milli\metre}$	\\
Slider width ($y$)		& $B$		& $\unit{6}{\milli\metre}$	\\
Number of blocks		& $N_x$		& $57$				\\
				& $N_z$		& $31$				\\
Slider mass			& $M$		& $\unit{75.6}{\gram}$		\\
Block mass			& $m$		& $M/(N_xN_z)$			\\
Young's modulus			& $E$		& $\unit{3}{\giga\pascal}$	\\
Bulk spring modulus		& $k$		& $3BE/4$		\\
Bulk spring length		& $l$		& $L/(N_x-1) = H/(N_z-1)$	\\
Damping {coefficient}		& $\eta$	& $\sqrt{0.1km}$		\\
Normal load			& $F_N$		& $\unit{1920}{\newton}$	\\
%Normal load asymmetry		& $\theta$	& $0$				\\
Elastic foundation modulus	& $k_f$		& $k/2$				\\
Driving spring modulus		& $K$		& $\unit{4}{\mega\newton\per\meter}$	\\
Driving height			& $h$		& $\unit{5}{\milli\metre}$	\\
Driving speed			& $V$		& $\unit{0.4}{\milli\metre\per\second}$	\\
\hline
Threshold force coefficient	& $\mu_s = f_\mathrm{thres}/f_N$	& $0.4$				\\
Slipping force coefficient	& $\mu_d = f_\mathrm{slip}/f_N$	& $0.17$			\\
Number of interface springs per block	& $N_s$		& $100$				\\
Interface spring stiffness	& {$k_{ij}$}	& $\sqrt{\unit{39.2}{\giga\newton\per\meter^2}f_{N,ij}}$ \\
%&	& {$\sqrt{\frac{w_{i}}{<w_i>_{N_x}}}\frac{K_I}{N_x N_s}$} \\
Slipping time mean		& $\bar t_R$	& $\unit{2}{\milli\second}$	\\
Slipping time standard deviation		& $\delta t_R$	& $\unit{0.6}{\milli\second}$	\\
Triggering region width		& $x_\mathrm{trigger}$ & $\unit{22.5}{\milli\metre}$ \\
Triggering region prestress	& {$\bar\tau_\mathrm{trigger}$}& 0.3\\
Time step duration		& $\Delta t$ & $\unit{2\e{-7}}{\second}$	\\
Extra damping coefficient	& $\alpha$	& $\eta/40$\\
\hline
\end{tabular}
\end{table}
\null
\vfill

% %%%%%%%%%%%%%%%%%%%%%%%%%%%%%%%
\clearpage
% %%%%%%%%%%%%%%%%%%%%%%%%%%%%%%%
\section{Supporting Methods}
This section supplements the model description found in the main text and its Methods section with detailed information on how we initialize the system and apply the boundary conditions. We also provide some additional information on the simplified model.

The slider is initialized with full normal load $F_N$ and no tangential load $F_T$ by gradually applying $F_N$ without allowing springs to break, a technicality required because the normal forces on the springs, $f_{Nij}$, start at zero and therefore springs, if allowed to, would break under any stretching. We distribute the load $F_N$ uniformly on the top blocks; apart from this we use the same non-frictional boundary conditions as in (\TromborgPRLrefnumber). The unique equilibrium is found through damped relaxation of typical duration $\unit{10}{\milli\second}$. After relaxation, we check that no spring is stretched beyond its strength and introduce the driving spring starting from zero applied driving force $F_T$. Then $F_T$, which acts on the block on the left side of the slider situated at height $h$ above the interface, through the driving spring, increases as the driving point moves to the right with speed $V$.

In the simplified model used for Fig. 3D, we disregard the microscopic state by using a single friction spring per block. Taking parameters from the microscopic reference model described in the main text, each block's spring now has a strength $\tau_\text{thres}=\mu_s p_i=N_s f_\text{thres}$. The stiffness $k_i$ of a block's friction spring equals the combined stiffness of the springs per block in the reference model. The force on block $i$ from its friction spring is $f_{Ti}=k_i (x_{is}-x_i)$, where $x_{is}$ is the attachment point of the spring to the track. Upon breaking, the spring becomes a slipping spring and its behaviour starts to differ from that of the springs in the microscopic model. We impose a slow slip by letting $x_{is}$ move with a velocity $v_\text{slow slip spring}$ for a time $t_\text{slow slip spring}=\bar t_R$. This process competes with the dynamic friction law where the spring trails the block with $x_{is}$ adjusted in every time step to ensure $\abs{f_{Ti}}\leq\mu_d p_i$, so that the spring attachment point moves with the highest of $v_\text{slow slip spring}$ and $v_{xi}$, the speed of the block in the $x$-direction. When the block motion reverses ($v_{xi}$ changes sign) the spring returns to the pinned state, but $x_{is}$ continues to move at $v_\text{slow slip spring}$ until $t_\text{slow slip spring}$ after the spring broke.

The systematic studies leading to Fig. 3C (inset, green), 3D and 4D were done with different normal forces and different initialization from the other simulations. The normal force boundary conditions on the top and bottom were exchanged: this simplifies the analysis by setting a constant normal force $p_i=F_N/N_x$ on all blocks $i$ at the interface. To maintain stability against global rotation, the top blocks interacted with an elastic ceiling with the same properties as the elastic foundation used in Ref. (\TromborgPRLrefnumber) and the other simulations presented here.

To obtain an initial state with a prescribed interfacial shear stress profile we turned the interface springs off during the initialisation. In their place we added to each bottom block the force corresponding to the shear stress to be prescribed. We also introduced the driving spring, but let $V=0$. During relaxation, the sample moved along the $x$-axis until the force in the driving spring balanced the net force from the interfacial shear stress. To get rid of oscillations more efficiently we added damping forces $-\alpha(\vec v_i)$ on the blocks' motion. After relaxation, the extra forces and the extra damping were turned off and the interfacial springs were introduced, with their attachment points $x_{ij}$ chosen such that the net force on each block was unchanged and the desired distribution of spring forces, $\phi(f_T)$, appeared. We then waited a few timesteps to ensure that the transition from pre- to post-relaxation involved no force discontinuities. Next, instead of driving the system with $V\neq­0$ until rupture was triggered, we started fronts by depinning simultaneously all springs for all blocks to the left of $x_\text{trigger}$. The shear stress in the triggering region has a strong influence on the rupture fronts; in order to compare results between simulations we used a constant value $\bar\tau_\text{trigger}$.

% %%%%%%%%%%%%%%%%%%%%%%%%%%%%%%%
\clearpage
% \section{Supporting Figures and Table}
% %%%%%%%%%%%%%%%%%%%%%%%%%%%%%%%
% \begin{figure}[t]
% \centering
% \includeordonotincludegraphics{}{./QuasiStatic/FIGSQuasiStatic}%
% %\includeordonotincludegraphics{}{./QuasiStatic/FIGSQuasiStatic_capital}%
% \caption{Quasi-static front propagation. Lowering the driving point to the interface ($h=0$, similar to 1D models) changes the dynamics of the system to a regime where rupture invades the interface quasi-statically on the timescale of the increase in $F_T$. \fpf{A} Macroscopic loading curve. \fpf{B} Fraction of pinned springs shown as in Fig.~2. \fpf{C{\rm,} D} To illustrate that quasi-static front propagation depends on the increase in $F_T$ we restart the simulation behind (\fpl{A{\rm,} B}) and set the driving speed $V=0$ at $t=\unit{0.59}{\second}$. The rupture propagation stops.}
% \end{figure}

\begin{figure}
\centering
%\includeordonotincludegraphics{width=\textwidth}{./StressProfilesFullSimulation/StressProfiles_f5000_capital}%
\includeordonotincludegraphics{width=\textwidth}{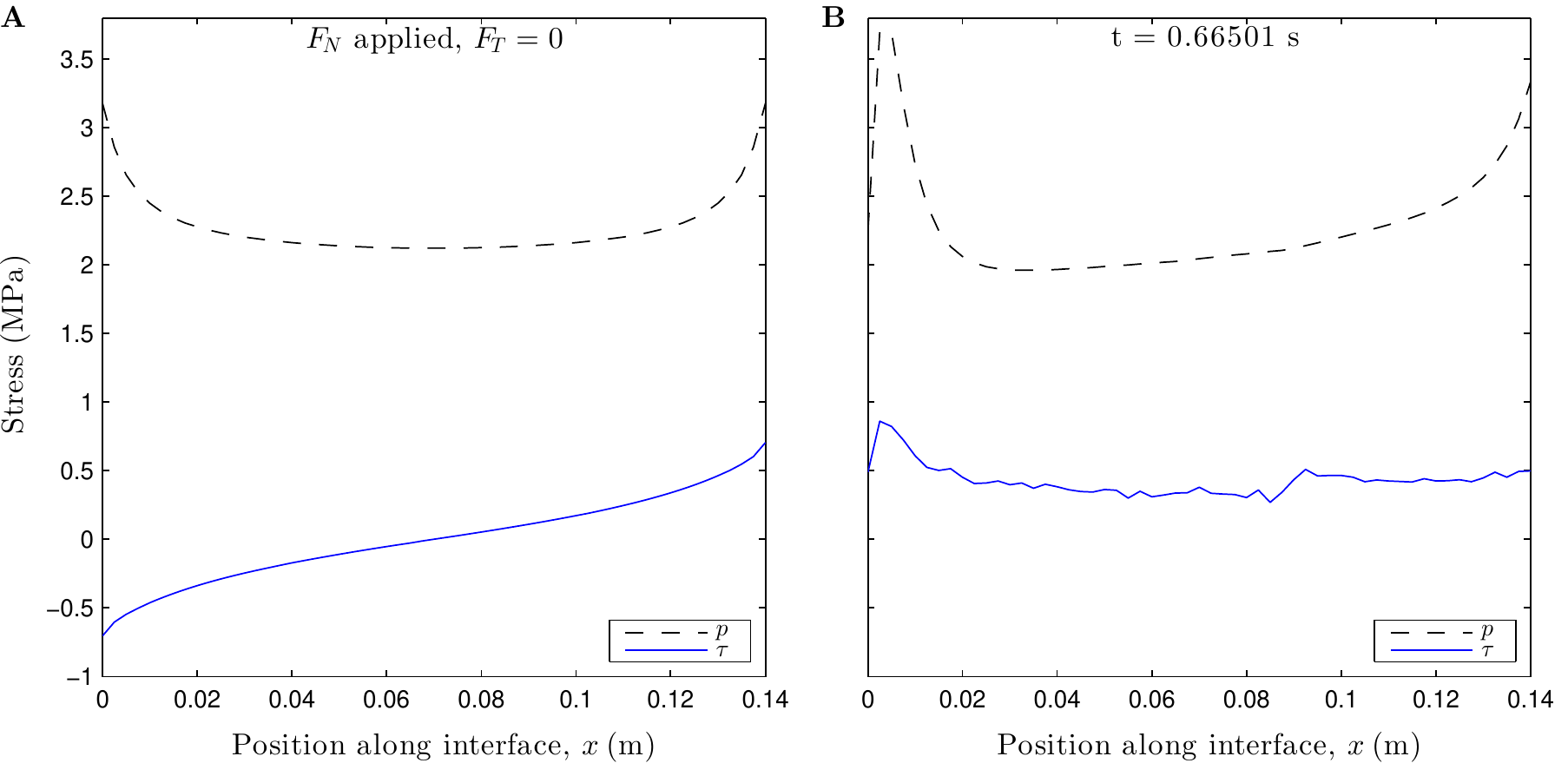}%
\caption{\Highlightchangesornot Spatial distributions of normal ($p$) and shear ($\tau$) stresses from the simulation behind events~I--III. \fpf{A} The initial state with full normal load $F_N$ and no tangential load $F_T$ (see Supporting Methods). The normal stress $p(x)$ is symmetric, with edge effects related to the flat punch geometry of the contact. The shear stress $\tau(x)$ is antisymmetric, due to Poisson expansion being restricted at the interface by friction. \fpf{B} The state just before event~I. The application of $F_T$ has modified both the shear stress profile and the normal stress profile (due to the friction-induced torque arising when $F_T$ is applied at a finite height $h$ above the interface).}
\end{figure}
% Referer til denne rett etter ``arising from macroscopic sample geometry and external loading''.

\begin{figure}
\centering
%\includeordonotincludegraphics{width=\textwidth}{./MesoscopicModel/slowFront_s7711_s771101}%
\includeordonotincludegraphics{width=\textwidth}{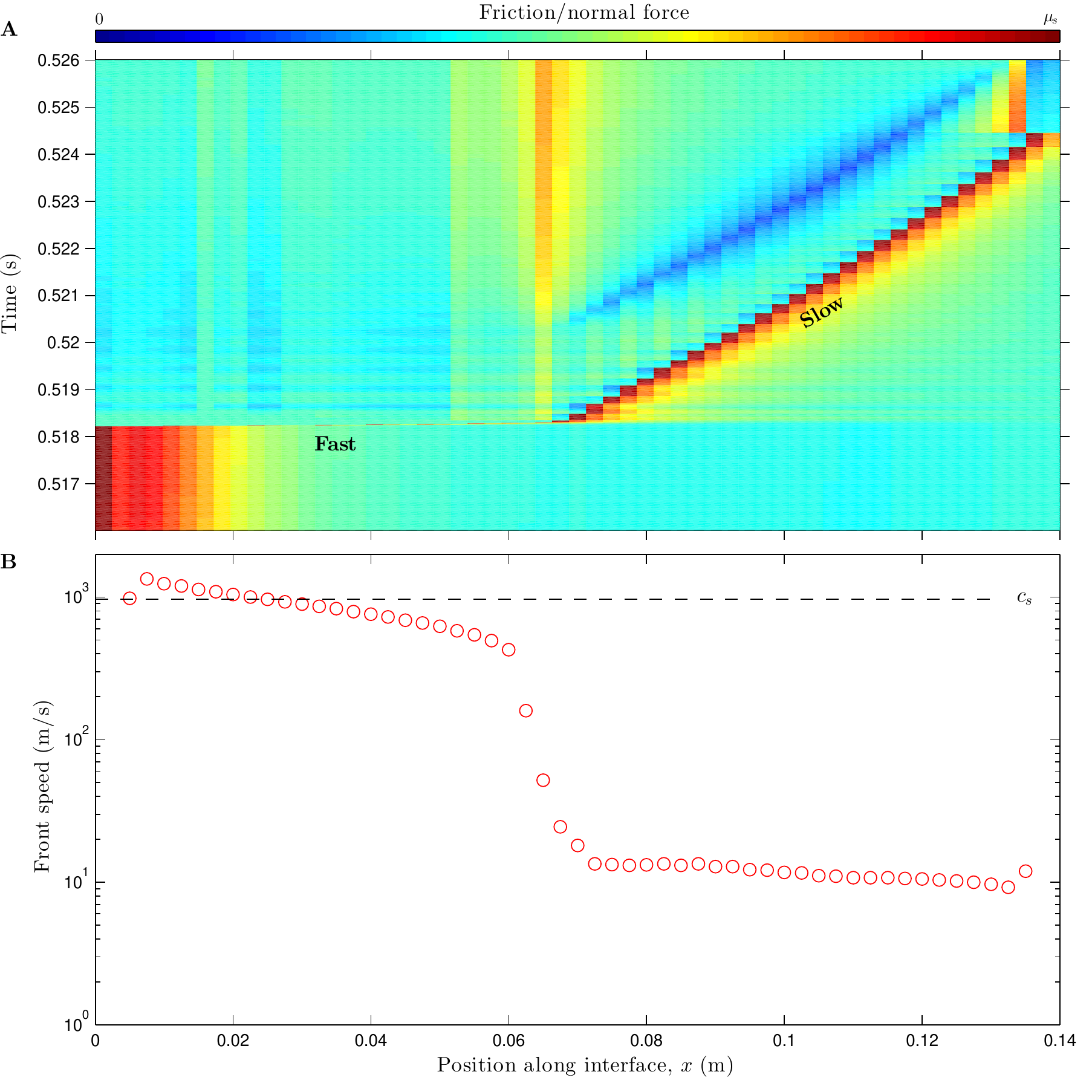}%
\caption{A fast-slow event arising spontaneously in a simulation using the simplified friction law with a single interface spring per block. The parameter $v_{\mathrm{slow}\,\mathrm{slip}\,\mathrm{spring}}=\unit{1.5}{\milli\meter\per\second}$ (Supporting Methods). \fpf{A} Spatio-temporal plot of the instantaneous friction to normal force ratio. \fpf{B} Rupture front speed $v_c$. Block rupture is defined to occur when the interface spring depins. Front speed is measured as in Fig.~2.} %(not shown)
\end{figure}

\begin{figure}
\centering
% \includeordonotincludegraphics{}{./TriggeringAndPropagationRegion/FIGS_triggeringAndPropagationRegion}%
\includeordonotincludegraphics{}{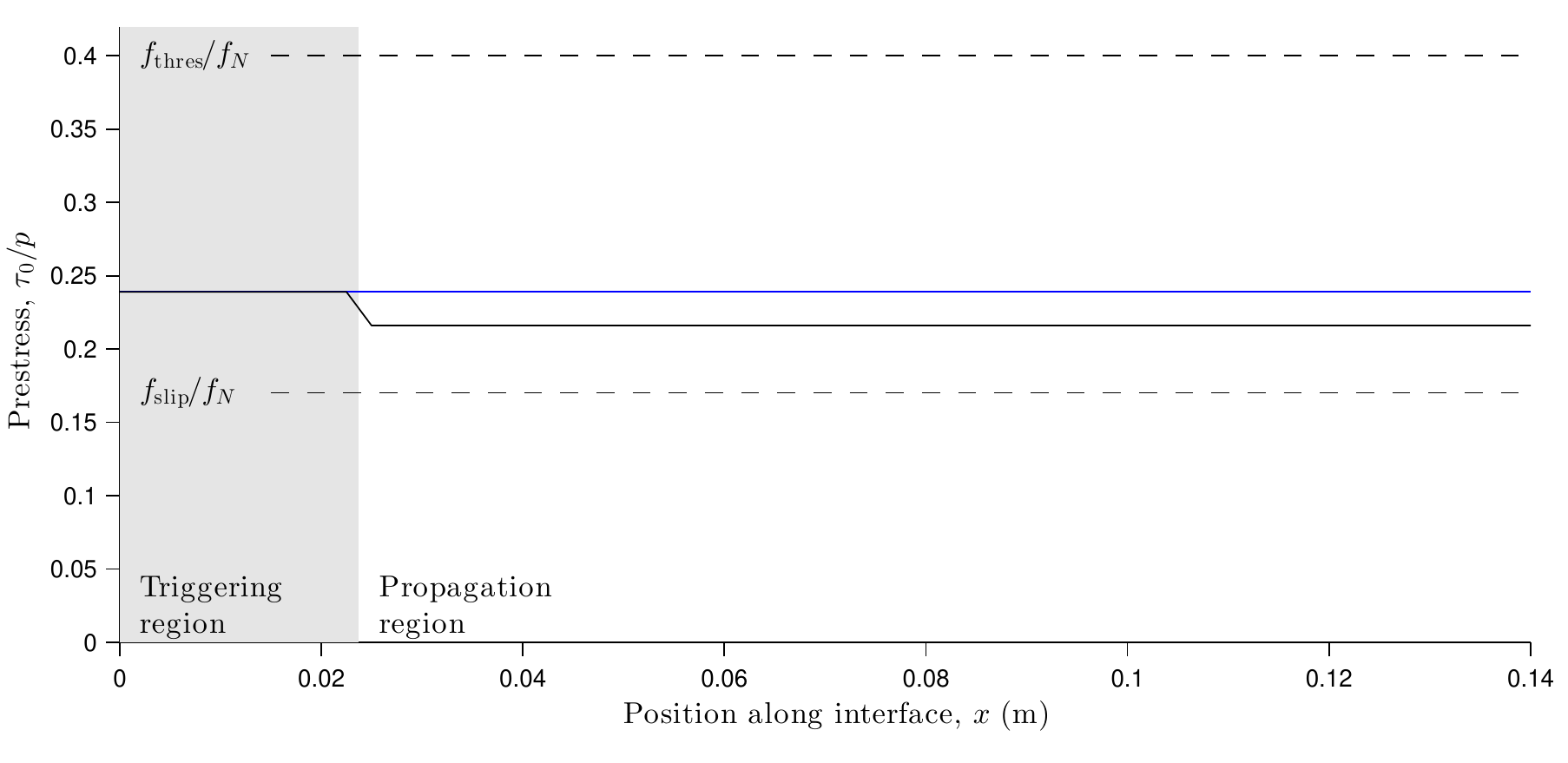}%
\caption{Spatial distributions of prestress $\tau_0/p$ for prepared states of the interface. In the front triggering region on the left, the prestress is $\bar\tau_0= \left(\frac{\tau_0}{p}-\frac{f_\mathrm{slip}}{f_N}\right)/\left(\frac{f_\mathrm{thres}}{f_N}-\frac{f_\mathrm{slip}}{f_N}\right)=0.3$ for all prepared states used in Fig.~3\fpt{D}, 4\fpt{C} and 4\fpt{D}. In the front propagation region the prestress is homogeneous along the interface, at a value varied between prepared states, here $\bar\tau_0=0.2$ (black) and $\bar\tau_0=0.3$ (blue). After initialization, all springs in the triggering region are depinned simultaneously. Initiating the events in this way, rather than by driving the system until rupture is triggered, ensures that the force drop / energy release in the triggering region remains the same between simulations.\\\\{\Highlightchangesornot The non-dimensional form $\bar\tau_0$ of the prestress represents the ratio between (i) the stress in excess of the stress obtained during sliding and (ii) the maximum dynamic stress drop that results from rupture. It is analogous to the so-called $S$ classically used in seismology and to the form defined in Ref. (\TromborgPRLrefnumber).}\label{fig:triggeringAndPropagationRegion}}
\end{figure}

\begin{figure}
\centering
%\includeordonotincludegraphics{}{./ModifyDistributionChangeFrontType/Fig_support_fastSlowFast_turned_fastOnly}%
\includeordonotincludegraphics{}{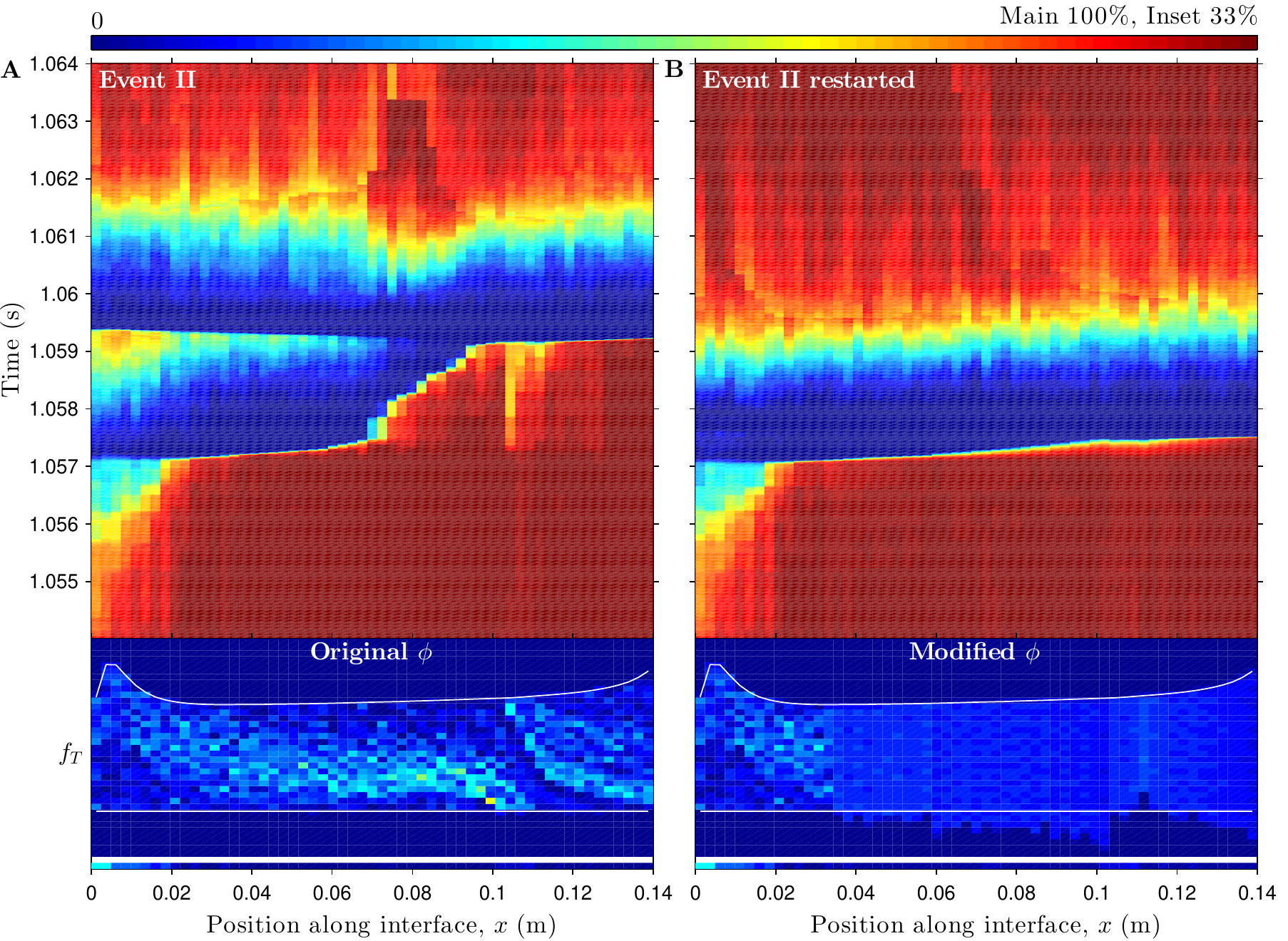}%
\caption{Microscopic force distributions significantly affect rupture fronts. The data in Fig.~4\fpt{A,B} repeated with spring force distributions shown in detail. \fpf{A} Event II shown as in Fig.~2. \fpf{B} Simulation behind \fpl{A} restarted at $\unit{1.054}{\second}$ with a wider distribution $\phi(f_T)$ of shear forces results in a fast-only event. Insets: For each block along the interface, a color coded histogram of $\phi(f_T)$ at $\unit{1.054}{\second}$. The vertical axis shows the force level in individual springs, which extends up to $f_\mathrm{thres}$. The level $f_\mathrm{thres}$ is shown by the upper white line; it is different for each block because it varies with normal force. The lower white line corresponds to $f_T=\unit{0}{\newton}$. The color denotes the fraction of springs found at each value of $f_T$ using an arbitrary bin width. This means that (apart from normalization) each vertical slice in the insets shows the same type of data as \figurename~\ref{fig:springStretchingToForceEvolution}\fpt{A}. Offset data: Fraction of slipping springs at $\unit{1.054}{\second}$.\\\\In order to isolate the effect of $\phi(f_T)$ on front propagation from the influence of front initiation and stress state we leave the loading zone on the left unmodified (it is the same in both insets) so that the restarted event begins like the original; we also let the modified $\phi(f_T)$ have the same mean value as the original $\phi(f_T)$ for all blocks. Thus, the stress state is the same and the only change is in the width of $\phi$.}
\end{figure}

\begin{figure}
\centering
% \includeordonotincludegraphics{}{./SpringStretching_forceCurve_musEff/FigS4b}%
% \includeordonotincludegraphics{}{./SpringStretching_forceCurve_musEff/FigS4b_capital}%
\includeordonotincludegraphics{}{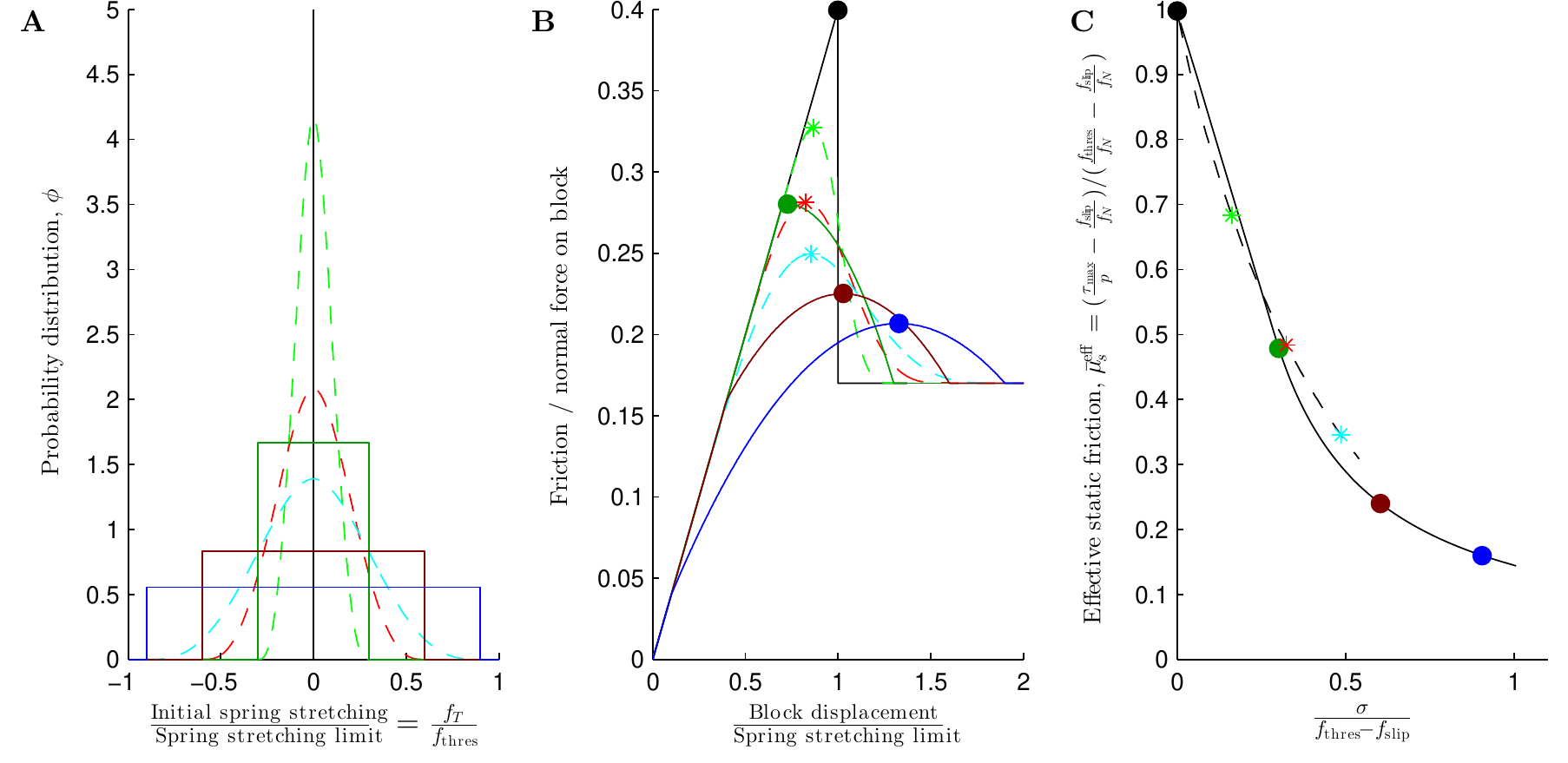}%
\caption{Dependence of effective static friction coefficient $\mu_s^\mathrm{eff} = \tau_\mathrm{max}/p$ on the distribution $\phi$ of forces in the interface springs. \fpf{A} Uniform and bell-shaped initial distributions of various widths. The bell-shaped distributions are polynomials with roots at $\pm a$ and the functional form $\phi(\xi) = 5/(4a)(1+3\abs{\xi/a})(1-\abs{\xi/a})^3, \xi\in[-a,a]$. \fpf{B} Friction to normal force ratio vs block displacement for a block having the spring force distributions in \fpl{A} (corresponding colours). Spring relaxation during slip is excluded from the calculation, a valid assumption when the passage of the rupture front is quick compared to the mean slipping time $\langle t_R\rangle$. Markers are located at maxima, which define $\tau_\mathrm{max}/p$. \fpf{C} Effective static friction coefficient vs $\sigma$, the standard deviation of $\phi$. Each marker takes its abscissa from the data in \fpl{A} and its ordinate from \fpl{B}. Drawn and dashed lines connect markers corresponding to uniform and bell-shaped $\phi$, respectively. They are used in Fig.~4\fpt{C} as reference lines for simulation data. For clarity, only a few of the $\phi$ that were used to determine the lines are shown on this figure.\\\\{\Highlightchangesornot The form $\bar\mu_s^\mathrm{eff}$ is analogous to $\bar\tau_0$ introduced in \figurename~\ref{fig:triggeringAndPropagationRegion}, and represents the ratio of (i) the dynamic stress drop from the effective static friction level to the stress obtained during sliding and (ii) the maximum value this stress drop can attain.}\label{fig:springStretchingToForceEvolution}}
\end{figure}

\begin{figure}
\centering
% \includeordonotincludegraphics{}{./LinearlyDecreasingDynamicFriction/NAS_and_phaseDiagram}%
% \includeordonotincludegraphics{}{./LinearlyDecreasingDynamicFriction/NAS_and_phaseDiagram_capital}%
\includeordonotincludegraphics{}{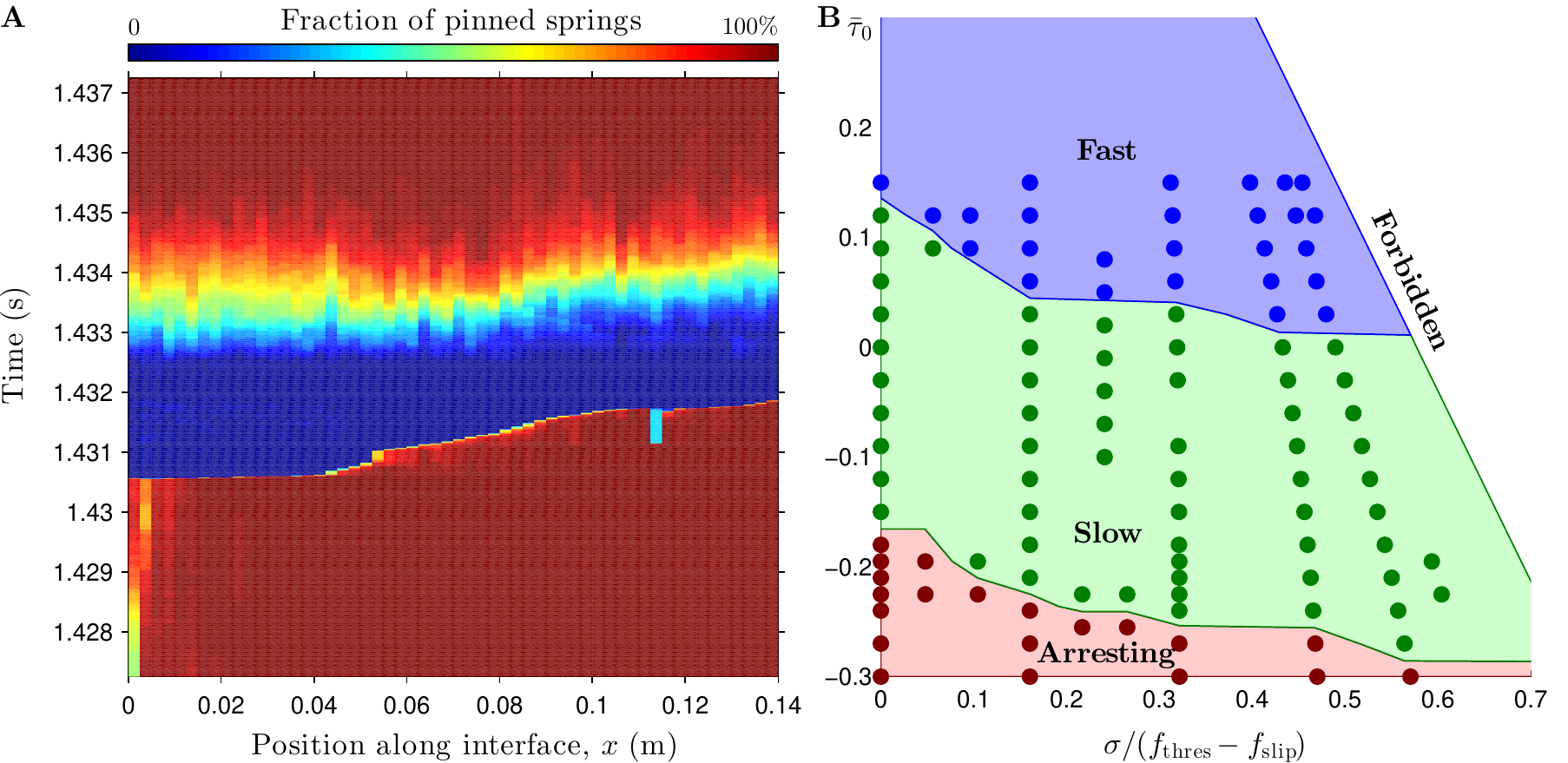}%
\caption{Results from an alternative model with a different individual behaviour of the interface springs, showing the robustness of our results against changes in the interfacial law. \fpf{A} A spontaneously arising fast-slow-fast event, analogous to that in Fig.~2\fpt{A}. \fpf{B} Observed front type for prepared interfaces, analogous to Fig.~4\fpt{D}.\\\\All data in this figure comes from simulations with a model in which the slipping force $f_\mathrm{slip}$ decreases linearly with time. This microscopic friction law modifies the slow slip mechanism with respect to the reference microscopic model. Until depinning, the springs have the same behaviour in both models. Upon entering the slipping state, the force from each spring on the block immediately drops to the level $f_\mathrm{slip}^0 = \mu_d f_{N,ij}$, as in the reference model. Then the slipping force decays linearly in time with a decay rate depending on the slipping time and the force level at repinning $f_\mathrm{new}$; that is, $f_\mathrm{slip} = f_\mathrm{slip}^0-(f_\mathrm{slip}^0-f_\mathrm{new})t_s/t_R$, with $t_s$ measured from the time the spring entered the slipping state. The slipping time distribution is the same as in the reference model. The decay in the friction coefficient enables a slow slip motion after the initial rapid slip, which allows this model to sustain slow fronts for a relatively wide range of initial spring stretching configurations and stress states, as seen in \fpl{B}.\\\\The parameters used with this model are the same as in the reference model (\tablename~\ref{tab:parameters}), except for $k_{ij}=\sqrt{\unit{54.1}{\giga\newton\per\meter^2}f_{N,ij}}$, $\mu_d=0.28$ and the new parameter $\mu_\mathrm{new}=0.7\mu_d \approx 0.2$ used to determine $f_\mathrm{new} = \mu_\mathrm{new}f_{N,ij}$.}
\end{figure}

% %%%%%%%%%%%%%%%%%%%%%%%%%%%%%%%
\clearpage
\section{Supporting Equations}
% %%%%%%%%%%%%%%%%%%%%%%%%%%%%%%%
\newcommand{\vcs}{v_{c,\mathrm{slow}}}
\newcommand{\vsslip}{v_{\mathrm{slow}\,\mathrm{slip}}}
\newcommand{\vsslipestimate}{v_\text{slow slip, estimate}}

{\Highlightchangesornot
\subsection{Estimate of \texorpdfstring{$\vsslip$}{v\_slowslip} from model parameters}
%The inset of Fig.~3\fpt{C} plots the slow slip speed $\vsslip$ measured in simulation data against an estimated $\vsslipestimate$ based on the simulation parameters. 
Here we provide the arguments behind the slow slip speed estimate used in the inset of Fig.~3\fpt{C}.

The mechanism for slow slip in our model is the force drop when slipping junctions relax and repin at zero force.
%are replaced by pinned junctions. The mechanism is active whenever such replacement occurs, but the slow slip it produces can be masked by other sources of slip, for example net motion of the entire slider. Here we consider the case of slow slip as a driver for slow fronts, but the argument can readily be extended to more general situations.
%Slip speed is block motion per unit time.
To determine the slip speed (block motion per unit time) associated with the relaxation of junctions, we identify the net slip caused by this change and the time over which the change happens. To do this, we assume that each time a junction relaxes, the block moves forward just enough to return to the force it was bearing just before relaxation. Thus, the force reduction $-f_\text{slip}$ in the junction is matched by an equivalent net force increase in the junctions that are already pinned and in the coupling to the neighbouring blocks. The effective stiffness of these interface and bulk terms depend on the fraction of junctions that are pinned and on the motion of the neighbouring blocks, respectively.

A careful look at the motion of blocks during their slow slipping regime proved that neighbouring blocks move essentially at the same slow slip speed. This means that the force changes on a block due to relative motion with respect to its neighbours remain negligible. We can therefore assume in this calculation that the only contribution to force restoring after junction relaxations is due to the pinned interfacial junctions.

We now assume that all junctions start in the slipping state.  
After the first junction relaxes and repins, the effective stiffness of the interface is just the stiffness of this single junction, $k_{ij}$. The force drop $-f_\text{slip}$ must be compensated by stretching this (now pinned) junction by moving the block a distance $\delta_1 = f_\text{slip}/k_{ij}$. For the second junction the force drop is again $-f_\text{slip}$, but the stiffness of the interface has increased to $2k_{ij}$ and the required slip is $\delta_2 = f_\text{slip}/(2k_{ij})$. If no junction would reach its breaking threshold $f_\text{thres}$ during the whole relaxation process, then we would find $x_\text{slow slip} = \sum_{j=1}^{N_s} \delta_j = \sum_{j=1}^{N_s} f_\text{slip}/(jk_{ij})$. Since the $k_{ij}$ are independent of $j$ this is just $\delta = f_\text{slip}/k_{ij}\sum_{n = 1}^{N_s}1/n$ with $n$ a dummy index. For $N_s=100$ used in the model, the sum evaluates to $5.2$. However, for $f_\text{thres}/f_\text{slip}=40/17$ used in the model, some junctions do break again before all slipping junctions relax. The force drop associated with the transition from pinned to slipping state is $f_\text{slip}-f_\text{thres}$, and acts in the same way as the force drop when junctions leave the slipping state. Taking this into account and evaluating the return to the pinned state more carefully, we find the net block slip to be $8.5f_\text{slip}/k_{ij}$.
%
%when the force in them exceeds $f_\text{thres}$, and with $f_\text{thres}/f_\text{slip}=40/17$ the first junctions to return to the pinned state will start slipping before the last junctions return to the pinned state. There is an associated force drop of $f_\text{thres}-f_\text{slip}$, acting in the same way as the force drop when junctions leave the slipping state. Taking this into account and evaluating the return to the pinned state more carefully the net block slip for $f_\text{thres}/f_\text{slip}=40/17$ and $N_s=100$ becomes $8.5N_sf_\text{slip}/k_{ij}$.

With the slipping time standard deviation $\delta t_R = 0.3\bar t_R$ used in the model, the time for all the junctions to return to the pinned state is found to be close to $2\langle t_R\rangle$.

Combining these slip distance and slip time values, and defining $\tau_\text{slip} = N_s f_\text{slip}$ and $k_i=N_s k_{ij}$, we use
\begin{align}
v_\text{slow slip, estimate} = 4.2 \frac{\tau_\text{slip}}{k_i\langle t_R\rangle}.
\end{align}
The inset of Fig.~3\fpt{C} plots the slow slip speed measured in the simulations against this estimate. The markers indicate when we have varied $\tau_\text{slip}$ ($\blacksquare$), $k_i$ ($\blacktriangleleft$) and $\langle t_R\rangle$ ($\blacktriangle$). $\bullet$ uses our reference parameters. The blue data is based on restarting event II. Because $\tau_\text{slip}$ and $k_i$ enter in the elastic state of the slider, only $\langle t_R\rangle$ could be varied for these simulations. The green data is for prepared homogeneous interfaces, where parameters can be varied freely.\footnote{When $f_\text{slip}$ is varied, $f_\text{thres}$ is varied proportionally in order to keep the fast-slow-fast nature of the rupture front. For the same reason, when $k_i$ is varied the prestress in the triggering region is also varied slightly.}

When slow slip speed is measured in the simulations based on event~II (blue data in inset of Fig.~3\fpt{C}), some deviation from $v_\text{slow slip, estimate}$ are observed. There are several possible reasons for such a deviation. First, the assumption of co-moving neighbours is only approximately correct (near the front tip the neighbours to the right are stuck until the rupture front passes them). The actual motion of the neighbours also depends on the stress state and the triggering of the event. Second, the assumption of force re-balancing every time a junction changes state is probably too strong.
}

\subsection{Scaling of \texorpdfstring{$\vcs$ with $\vsslip$}{v\_c,slow with v\_slowslip}}
The data collapse in Fig.~3\fpt{D} is obtained by plotting the slow rupture speed $\vcs$ against the quantity $\vsslip k_il_0/(\tau_\mathrm{thres}-\tau^0)$, where $\vsslip$ is the slow slip speed, $k_i$ is the stiffness of the connection between a block and the interface (a single spring in the simplified model and a parallel connection in the reference microscopic model), $l_0$ is a characteristic length, $\tau_\mathrm{thres}=N_sf_\mathrm{thres}$ is the maximum shear strength of a block and $\tau^0$ is the shear force in the propagation region before the event is started. In this section we provide a crude argument for this scaling.

When a region of initially homogeneous prestress is being stressed further by block motion on the left, the decaying shear force profile can be written on the form
\begin{align}
\tau(x) = Af\left(\frac{x-x_0}{l_0}\right)+\tau^0,
\end{align}
where $A$ is an amplitude and $f()$ is a function that has magnitude $1$ at $x=x_0$ and decays over a characteristic length $l_0$ that depends on the bulk to interfacial stiffness ratio $k/k_i$. The function $f()$ is known in 1D (Ref.~(\AmundsenTribolLettrefnumber), equation~(46)); in 2D it can be measured in an elastostatic model, but its exact form is not required for the present argument.%ref to Amundsen 2012

In a static situation $\tau(x)$ is balanced by the friction forces in the interfacial springs. Ignoring the width of the spring force distribution, the block at $x_0$ is at its static friction threshold when the force on it from its neighboring blocks is $\tau(x_0)=\tau_\mathrm{thres}$, which gives $A=(\tau_\mathrm{thres}-\tau^0)$. The next block to the right, at position $x=x_0+\dx$, then has
\begin{align}
\tau(x_0+\dx) = Af[x_0+\dx]+\tau^0.
\end{align}
Here we have used the short-hand notation $f[x]=f\big((x-x_0)/l_0\big)$.

As the front tip moves from the block at $x_0$ to the block at $x_0+\dx$, the force on this block from its neighbours increases to $\tau'(x_0+\dx)=\tau_\mathrm{thres}$. It will be useful to rewrite this as $\tau'(x_0+\dx)=\tau_\mathrm{thres}=\tau(x_0)=Af[x_0]+\tau^0$. The change in force on the block at $x_0+\dx$ is
\begin{align}
\Delta\tau(x_0+\dx) &= \tau'(x_0+\dx)-\tau(x_0+\dx)\\
&= A(f[x_0]-f[x_0+\dx]).
\end{align}
Assuming a corresponding change in the friction force allows us to relate the force change to a displacement of the block, namely
\begin{align}
\Delta u(x_0+\dx) = \frac{\Delta\tau(x_0+\dx)}{k_i}.
\end{align}
In the next step we will need the displacement of the block at $x_0$ during the same time interval. As the blocks are at closely spaced points in a deforming elastic medium we will assume $\Delta u(x_0) = \Delta u(x_0+\dx)(1+O(\dx)) \approx \Delta u(x_0+\dx)$.

Now we make the approximation that after breaking, the blocks move at a constant speed $\vsslip$. It follows that the time it takes from when the block at $x_0$ breaks and until when the block at $x_0+\dx$ breaks is
\begin{align}
\mathrm{d}t = \frac{\Delta u(x_0)}{\vsslip}.
\end{align}
During this time the front tip has moved the distance $\dx$ from one block to the next, and the front speed is
\begin{align}
\vcs = \frac{\dx}{\mathrm dt} &= \vsslip\frac{\dx}{\Delta u(x_0)}\\
&= \vsslip\frac{k_i}{\tau_\mathrm{thres}-\tau^0}\frac{dx}{f[x_0]-f[x_0+\dx]}.
\end{align}
Here we recognize an approximation to the spatial derivative of the unknown function $f()$, evaluated at $x=x_0$.
% This derivative can be written on the form $\mathrm df/\dx = -g[x]/l_0[k/k_i]$, with $g[x]$ dimensionless because we have extracted the length scale as $1/l_0$, which is the derivative of the kernel of $f()$. Using this form we arrive at
% \begin{align}
% \vcs = \vsslip\frac{k_i}{\mu_sp-\tau^0}\frac{l_0[k/k_i]}{g[x_0]}.
% \end{align}
We will use the chain rule to separate the non-dimensional and dimensional parts of this derivative, and therefore we define $X(x)=(x-x_0)/l_0$ so that $f[x]=f(X(x))$. With this notation,
\begin{align}
\frac{\mathrm{d}f}{\mathrm{d}x} &= \frac{\mathrm{d}f}{\mathrm{d}X}\frac{\mathrm{d}X}{\mathrm{d}x}
= \frac{\mathrm{d}f}{\mathrm{d}X}\frac{1}{l_0}.
\end{align}
We arrive at
\begin{align}
\vcs = \vsslip\frac{k_i}{\tau_\mathrm{thres}-\tau^0}\frac{l_0}{\left(-\frac{\mathrm{d}f}{\mathrm{d}X}\right)_{x=x_0}}.
\end{align}

This argument provides a rationale for the linear relationship observed in Fig.~3\fpt{D}, but with the function $f()$ unknown we are not able to predict the value of the coefficient of proportionality. From the shear force profiles we estimate the decay length $l_0=\unit{7}{\milli\meter}$, a value shared between simulations because we keep $k/k_i$ the same, and rescale $\vsslip$ with $k_il_0/(\tau_\mathrm{thres}-\tau^0)$. Note that in the model, $\tau_\mathrm{thres} = N_sf_\mathrm{thres} = \mu_sp$, with $\mu_s$ the threshold force coefficient and $p$ the normal force on the block, which means that the normal force enters in the scaling.

\end{document}